\documentclass[12pt,letterpaper]{article}%
\usepackage[T2A]{fontenc}
\usepackage[cp1251]{inputenc}
\usepackage{amsmath}
\usepackage{amsfonts}
\usepackage{amssymb}
\usepackage{graphicx}
\usepackage{cite}%
\newcommand\PACS[1]{\vskip-2.75pc \begin{center}\parbox{.8\textwidth}{\small\bf PACS numbers: \rm #1 \hfill} \end{center}\vskip4pt}%

\begin{document}
\title{Ray-based description of mode coupling by sound speed
fluctuations in the ocean}
\author{A.L. Virovlyansky\\
Institute of Applied Physics, Russian Academy of Science,\\
46, Ulyanova St., Nizhny Novgorod, Russia, 603950 \\
 viro@hydro.appl.sci-nnov.ru
}

\maketitle

\begin{abstract}
A traditional approach to the analysis of mode coupling in a fluctuating
underwater waveguide is based on solving the system of coupled equations for
the second statistical moments of mode amplitudes derived in the Markov
approximation [D.B. Creamer, J.Acoust. Soc. Am. 99, 2825--2838 (1996)]. In the
present work an alternative approach is considered. It is based on an analytic
solution of the mode coupling equation derived in the high frequency
approximation [A.L. Virovlyanskii and A.G. Kosterin, Sov. Phys. Acoust. 35,
138-142 (1987)]. This solution representing the mode amplitude as a sum of
contributions from two geometrical rays is convenient for statistical
averaging. It allows one to easily derive analytical expressions for any
statistical moments of mode amplitudes. The applicability of this approach is
demonstrated by comparing its predictions for a deep water acoustic waveguide
with results of full wave numerical simulation carried out using the method of
wide angle parabolic equation.


\end{abstract}

\newpage

\section{Introduction \label{sec:introduction}}

An environmental model representing an unperturbed range-independent waveguide
with weak sound speed fluctuations is widely used in solving different
problems of ocean acoustics \cite{FDMWZ79,BL2003}. The sound field in this
model can be decomposed into a sum of normal modes of the unperturbed
waveguide with complex amplitudes $a_{m}$ being random functions of range $r$.
Mode coupling by the sound speed fluctuations is quantitatively described by
statistical moments of mode amplitudes $\left\langle a_{m}a_{n}^{\ast
}\right\rangle $ where the asterisk denotes complex conjugation and the
angular brackets denote statistical averaging. A traditional approach to the
analysis of the modal structure of the sound field in a fluctuating waveguide
is based on solving the transport equations for moments $\left\langle
a_{m}a_{n}^{\ast}\right\rangle $
\cite{Papani77,DT1,DT2,Creamer1996,Sazontov2002,Colosi2009}.

Complete system of transport equations derived in the Markov approximation
includes $M^{2}$ linear differential equations, where $M$ is the number of
propagating modes \cite{Creamer1996,Colosi2009}. A truncated version of this
system including only $M$ equations for the mean mode intensities
$\left\langle \left\vert a_{m}\right\vert ^{2}\right\rangle $ -- they are
called the master equations -- was derived in Ref. \cite{DT1} under assumption
that the cross-mode coherences, that is, the moments $\left\langle a_{m}%
a_{n}^{\ast}\right\rangle $ with $m\neq n$, are negligible.

In Ref. \cite{Colosi2009} all the statistical moments $\left\langle a_{m}%
a_{n}^{\ast}\right\rangle $ were calculated numerically for a deep water
acoustic waveguide with sound speed fluctuations induced by random internal
waves. It turned out that the values of $\left\vert \left\langle a_{m}%
a_{n}^{\ast}\right\rangle \right\vert $ for close, but not equal, $m$ and $n$
are of the same order of magnitude as $\left\langle \left\vert a_{m}%
\right\vert ^{2}\right\rangle ^{1/2}\left\langle \left\vert a_{n}\right\vert
^{2}\right\rangle ^{1/2}$ and the mean intensities and cross-mode coherences
'evolve over similar range scales'. This contradicts the assumption about the
smallness of the cross-mode coherences made when deriving the master
equations. In spite of this fact, the numerical simulation demonstrated that
the solutions of master equations provide good estimates of mean mode
intensities $\left\langle \left\vert a_{m}(r)\right\vert ^{2}\right\rangle $.
However, it should be noticed that these solutions are smooth functions of
range and they do not predict small oscillations of $\left\langle \left\vert
a_{m}(r)\right\vert ^{2}\right\rangle $ with range found by solving the complete system
of all the $M^{2}$ equations. Similar
results were obtained in Refs. \cite{Colosi2012,Colosi2014} for a shallow
water waveguide.

In the present paper we consider an alternative approach to examining the mode
coupling. It was derived in Ref. \cite{V87} (see also Refs.
\cite{V89,V89a,V91b,V97}) in the study of ray-mode relations in a waveguide
with weak sound-speed fluctuations. This approach is based on a surprisingly
simple analytical solution of the mode coupling equation obtained in the high
frequency approximation. It expresses the mode amplitude through parameters of
two geometrical rays which we call the mode rays. This estimate of $a_{m}$ is
an analog of the well-known formula of geometrical optics describing the
variation of complex ray amplitude in the presence of weak inhomogeneities of
refractive index. The ray-based estimate of mode amplitude is convenient for
statistical averaging. Analytical expressions for any statistical moments of
mode amplitudes, including joint moments of amplitudes at different
frequencies, are readily follow from this formula.

We also consider a simplified expression for the mean mode intensity obtained
by averaging the parameters of the ray-based estimate of $\left\langle
\left\vert a_{m}\right\vert ^{2}\right\rangle $ over the ray cycle length. It
is shown that in the high frequency limit this expression, describing a
smoothed range-dependence of the mean mode intensity, satisfies the master
equations. This fact agrees with numerical result of Refs.
\cite{Colosi2009,Colosi2012,Colosi2014}.

A weak point of our analytical approach is insufficient knowledge
about the limits of its applicability. So far, its predictions
have never been compared to results of full wave simulation. In
the present paper such a comparison is
made for an environmental model similar to that used in Ref. \cite{Colosi2009}%
. Sound field excited by a points source at a frequency of 100 Hz was
calculated numerically using the method of wide angle parabolic equation in
360 realizations of the fluctuating waveguide. Mode amplitudes were found by
projecting computed wave fields onto eigenfuntions of the unperturbed
waveguide. Parabolic equation based (pe-based) estimates obtained in this way
are compared with the ray-based estimates found by evaluating mode amplitudes
in the same realizations of the waveguide using our approximate analytical
solution of the mode coupling equation. Estimates of statistical moments were
computed by the Monte Carlo method, that is, by averaging the products of mode
amplitudes over all the realizations of random waveguide. The comparison has
demonstrated a good agreement between the pe-based and ray-based results.

Numerical simulation have shown that the ray-based approach
properly describes not only the smoothed range-dependencies of
statistical moments of mode amplitudes but the small oscillations
of these moments (''missed'' by the master equations), as well. 
These oscillations were predicted in
Ref. \cite{V89} where it was shown that the jump-like variations
of the mode amplitude and its statistical moments occur in the
neighborhood of the upper turning points of the mode rays.

As in Refs. \cite{DT1,Creamer1996,Colosi2009,Colosi2012,Colosi2014}, we
neglect the horizontal refraction of sound waves and consider a
two-dimensional environmental model. Out-of-plane wave scattering was taken
into account in Refs. \cite{Voronovich06,Voronovich09}.

The organization of this paper is as follows.

Analytical relation expressing the mode amplitude through parameters of two
ray paths is presented in Sec. \ref{sec:ray-mode}. Section \ref{sec:moments}
gives analytical expressions for a few statistical moments of mode amplitudes
derived using this relation. In Sec. \ref{sec:master}, it is shown that in the
limit of high frequency the ray-based estimate of the mean mode intensity
$\left\langle \left\vert a_{m}\right\vert ^{2}\right\rangle $ whose parameters
are smoothed over the ray cycle length satisfies the master equations. Section
\ref{sec:numeric} presents results of numerical simulation in a deep water
waveguide with sound speed fluctuations induced by random internal waves. It
is demonstrated that the predictions of our ray-based approach agree with the
results of simulation carried out using the method of wide angle
parabolic equation. In Sec. \ref{sec:conclusion}, the results of this work are summarized.

\section{Analytical description of mode amplitudes in the presence of weak
sound speed fluctuations \label{sec:ray-mode}}

In this section we present a simple analytical approach derived in
Refs. \cite{V87,V91b,V97} for a ray-based description of mode
amplitudes in a waveguide with weak large scale sound speed
fluctuations. It is assumed that the wave field is excited by a
point source.

\subsection{Mode representation of the wave field \label{sub:WKB}}

Consider a two dimensional model of underwater sound channel with the sound
speed field $c(r,z)=\bar{c}(z)+\delta c\left(  r,z\right)  $, where $r$ is the
distance, $z$ is the depth, $\bar{c}(z)$ is the unperturbed sound speed
profile, and $\delta c\left(  r,z\right)  $ is the weak range-dependent
perturbation. The refractive index is $\nu\left(  r,z\right)  =c_{0}/c(r,z)$,
where $c_{0}$ is the reference sound speed. We assume that $\left\vert
c\left(  r,z\right)  -c_{0}\right\vert \ll c_{0}$. Due to the weakness of perturbation%

\[
\nu\left(  r,z\right)  =\bar{\nu}\left(  z\right)  +\delta\nu\left(
r,z\right)  ,
\]
where
\[
\bar{\nu}\left(  z\right)  =\frac{c_{0}}{\bar{c}\left(  z\right)  }%
,\;\delta\nu\left(  r,z\right)  =-\frac{c_{0}}{\bar{c}^{2}\left(  z\right)
}\delta c\left(  r,z\right)  .
\]

We assume that the perturbation $\delta c\left(  r,z\right)  $ is a zero mean
Gaussian random field with the correlation function
\begin{equation}
\left\langle \delta c\left(  r_{1},z_{1}\right)  \delta c\left(  r_{2}%
,z_{2}\right)  \right\rangle =K\left(  \xi,\zeta,Z\right)  , \label{K-def}%
\end{equation}
where%
\[
\xi=r_{1}-r_{2},\;\zeta=z_{1}-z_{2},\;Z=\left(  z_{1}+z_{2}\right)  /2.
\]
Note that%
\begin{equation}
K\left(  \xi,\zeta,Z\right)  =K\left(  -\xi,\zeta,Z\right)  . \label{K-sym}%
\end{equation}
Characteristic scales of correlation function $K$ along the coordinates $\xi$
and $\zeta$ denote $\Delta_{\xi}$ and $\Delta_{\zeta}$, respectively.

The acoustic pressure field $u\left(  r,z\right)  $ at a carrier frequency $f$
can be expressed as
\begin{equation}
u\left(  r,z\right)  =\sum_{m=1}^{M}\sqrt{\frac{2\pi i}{k_{m}r}}a_{m}\left(
r\right)  \varphi_{m}\left(  z\right)  e^{ik_{m}r}, \label{u-mode}%
\end{equation}
where $k_{m}$ and $\varphi_{m}\left(  z\right)  $ are eigenvalues and
eigenfunctions of the Sturm-Liouville problem in the unperturbed
(range-independent) waveguide, respectively \cite{BL2003,JKPS2011}. For
simplicity, it is assumed that the sum (\ref{u-mode}) includes only those
modes whose turning points are located within the water bulk. This assumption
will simplify the use of the WKB approximation for description of $k_{m}$ and
$\varphi_{m}\left(  z\right)  $.

In what follows we will consider the wave field excited by a point source set
at $r=0$ and $z=z_{0}$. In this case,
$ a_{m}\left(  0\right)  =\varphi_{m}\left(  z_{0}\right) $.

In the WKB approximation the eigenvalues can be presented as $k_{m}=kh_{m}$,
where $k=2\pi f/c_{0}$ is a reference wavenumber and $h_{m}$ is determined by
the quantization rule \cite{BL2003}%
\begin{equation}
k\int_{z_{\min}}^{z_{\max}}dz~\sqrt{\bar{\nu}^{2}(z)-h_{m}^{2}}=\pi\left(
m-1/2\right)  \label{quantization}%
\end{equation}
with $z_{\min}$ and $z_{\max}$ being the mode turning depths. In this
approximation the $m$-th mode is associated with a ray path whose grazing
angle at depth $z$, $\theta_{m}\left(  z\right)  $, is determined by the
relation
$h_{m}=\bar{\nu}\left(  z\right)  \cos\theta_{m}\left(  z\right)$.

The cycle length (period) of this ray path is given by%
\begin{equation}
D_{m}=2h_{m}\int_{z_{\min}}^{z_{\max}}\frac{dz}{\sqrt{\bar{\nu}^{2}%
(z)-h_{m}^{2}}}=2\int_{z_{\min}}^{z_{\max}}\frac{dz}{\tan\left(  \theta
_{m}\left(  z\right)  \right)  }~\text{.} \label{Dm}%
\end{equation}

A 'differential' form of the quantization rule follows from Eqs.
(\ref{quantization}) and (\ref{Dm}) as%
\begin{equation}
\frac{dh_{m}}{dm}=-\frac{2\pi}{kD_{m}}. \label{quantization-diff}%
\end{equation}

The eigenfunction $\varphi_{m}\left(  z\right)  $ in the WKB approximation
\cite{BL2003} can be presented in the form%
\begin{equation}
\varphi_{m}(z)=\varphi_{m}^{+}(z)+\varphi_{m}^{-}(z), \label{phi-sum}%
\end{equation}
where%
\begin{equation}
\varphi_{m}^{\pm}\left(  z\right)  =q_{m}\left(  z\right)  e^{\pm i\left[
kg_{m}\left(  z\right)  -\pi/4\right]  }, \label{phim-pm}%
\end{equation}%
\begin{equation}
q_{m}\left(  z\right)  =\frac{h_{m}^{1/2}}{\left[  \bar{\nu}^{2}(z)-h_{m}%
^{2}\right]  ^{1/4}D_{m}^{1/2}}=\frac{1}{\left[  D_{m}\tan\theta_{m}\left(
z\right)  \right]  ^{1/2}}, \label{Qm}%
\end{equation}%
\begin{equation}
g_{m}\left(  z\right)  =\int_{z_{\min}}^{z}dz~\sqrt{\bar{\nu}^{2}(z)-h_{m}%
^{2}}. \label{Gm}%
\end{equation}
Functions $\varphi_{m}^{\pm}\left(  z\right)  $ represent two quasi-plane
waves called the Brillouin waves.

Note a useful relation following from Eqs. (\ref{quantization}) and (\ref{Gm})%
\begin{equation}
k\frac{\partial g_{m}\left(  z\right)  }{\partial m}=\frac{2\pi}{D_{m}}%
\int_{z_{\min}}^{z}\frac{dz}{\tan\left(  \theta_{m}\left(  z\right)  \right)
}. \label{dGm-dm}%
\end{equation}

\subsection{Geometrical optics for modes \label{sub:geom}}

Within the framework of standard geometrical optics, the influence of a weak
sound speed perturbation $\delta c$ with spatial scales significantly
exceeding the wavelength can be accounted for using a well-known approximate
formula. If in the unperturbed medium the contribution of a sound ray to the
total field $u$ is $A\exp\left(  ikS\right)  $, where $A$ and $S$ are the ray
amplitude and eikonal, respectively, then in the presence of perturbation its
contribution becomes \cite{FDMWZ79,BL2003}
\begin{equation}
u=Ae^{ik(S+X)}, \label{A-ray}%
\end{equation}
where%
\begin{equation}
X=\int_{\Gamma}\delta\nu~ds, \label{X-geom}%
\end{equation}
$ds$ is the arc length and the integration goes over the unperturbed ray path
$\Gamma$. Although this formula is valid only at relatively short ranges it is
widely used in the ocean acoustics \cite{FDMWZ79}. In particular, it is used
in solving the inverse problem in the classical scheme of ocean acoustic
tomography \cite{MW79}.

In Ref. \cite{V87} (see also Refs. \cite{V89,V89a,V91b,V97}) it is shown that
there exists a close analog of Eq. (\ref{A-ray}) for normal modes. The point
is that the $m$-th mode constructively interferes (adds in phase) with
neighboring modes along the trajectories of two unperturbed rays leaving the
source at launch angles $\pm\theta_{m}\left(  z_{0}\right)  $ which are equal
to grazing angles of the Brillouin waves $\varphi_{m}^{\pm}\left(  z\right)  $
at the source depth. As in Refs. \cite{V87,V89,V89a,V91b,V97}, we shall call
these rays the mode rays and denote their trajectories $z_{m}^{\pm}\left(
r\right)  $. Note that for a given $m$ the angle $\theta_{m}\left(
z_{0}\right)  $ is a function of the carrier frequency $f$. This makes the
trajectories $z_{m}^{\pm}\left(  r\right)  $ frequency dependent. Denote the
grazing angles of the ray paths $z_{m}^{\pm}\left(  r\right)  $ at range $r$
by $\chi_{m}^{\pm}(r)$, so that $dz_{m}^{\pm}(r)/dr=\tan\chi_{m}^{\pm}(r)$.
Both mode rays have the same cycle length given by Eq. (\ref{Dm}). Examples of
mode rays are shown in Fig. 1. It graphs trajectories
$z_{m}^{\pm}(r)$ for the 36-th mode in the canonical sound speed profile at a
carrier frequency of 100 Hz.

In the presence of perturbation $\delta c\left(  r,z\right)  $ the mode
amplitude is expressed by the approximate formula%
\begin{equation}
a_{m}\left(  r\right)  =\varphi_{m}^{+}(z_{0})e^{ikX_{m}^{+}\left(  r\right)
}+\varphi_{m}^{-}(z_{0})e^{ikX_{m}^{-}\left(  r\right)  }, \label{am-ray}%
\end{equation}
with%
\begin{equation}
X_{m}^{\pm}=\int_{\Gamma_{m}^{\pm}}ds~\delta\nu, \label{Xmp-def}%
\end{equation}
where the integration goes along the trajectories of mode rays $\Gamma
_{m}^{\pm}$ \cite{V87,V97}. Equation (\ref{Xmp-def}) can be written in the
form%
\begin{equation}
X_{m}^{\pm}\left(  r\right)  =\int_{0}^{r}\frac{dr^{\prime}}{\cos\chi_{m}%
^{\pm}}\delta\nu\left(  r,z_{m}^{\pm}\left(  r\right)  \right)  .~
\label{Xmp1}%
\end{equation}

\begin{figure}[ptb]
\begin{center}
\includegraphics[
height=2.4094in,
width=5.975in
]{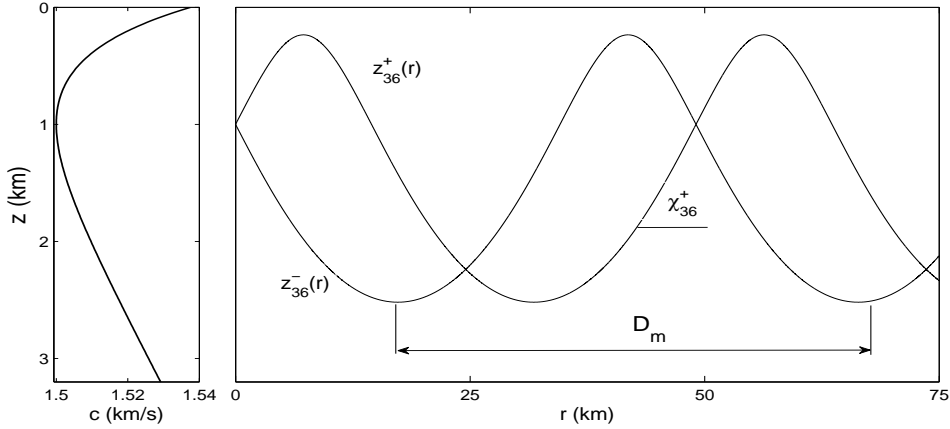}
\end{center}
\caption{ Left panel: canonical sound speed profile. Right panel: trajectories
of the mode rays of the 36-th mode at a carrier frequency of 100 Hz.}%
\label{fig_mode_ray}%
\end{figure}

Formula (\ref{am-ray}) is derived under the same assumptions as its prototype
for the ray amplitude (\ref{A-ray}). The simplest derivation of Eq.
(\ref{am-ray}) consists in projecting the ray representation of the wave field
onto eigenfunctions of the unperturbed waveguide with the
evaluation of arising integrals using the stationary phase technique
\cite{V91b,V97}. Therefore Eq. (\ref{am-ray}) should have approximately the
same range of applicability as Eq. (\ref{A-ray}).

A more accurate and general expression for the mode amplitude can be derived
proceeding from the ray representation of the wave field in a range-dependent
waveguide \cite{V99c,V2005a,Vbook2010}. Besides, Eq. (\ref{am-ray}) can be
generalized in a different direction. In Ref. \cite{V87,V91b,V97},\ a more
general version of this formula was derived which can be used for description
of wave diffraction by sound speed fluctuations. In these works, the notion of
Fresnel zones for modes is introduced which is analogous to the usual Fresnel
zones introduced for rays.

In the present paper the indicated generalizations are not considered. All our
subsequent analysis is based on formula (\ref{am-ray}).

\section{Statistical moments of mode amplitudes \label{sec:moments}}

Since $X_{m}^{\pm}\left(  r\right)  $ are zero mean Gaussian random functions,
an analytical expression for any statistical moment of mode amplitudes is
readily derived from Eq. (\ref{am-ray}) using the well-known formula
$\left\langle e^{i\alpha}\right\rangle =-e^{-\left\langle \alpha
^{2}\right\rangle /2}$ for a zero mean Gaussian random variable $\alpha$. The
mean value (coherent component) of the mode amplitude $a_{m}$ is%

\begin{equation}
\left\langle a_{m}\right\rangle =\Phi_{m}^{+}e^{-\frac{k^{2}}{2}\left\langle
\left(  X_{m}^{+}\right)  ^{2}\right\rangle }+\Phi_{m}^{-}e^{-\frac{k^{2}}%
{2}\left\langle \left(  X_{m}^{-}\right)  ^{2}\right\rangle }, \label{am-mean}%
\end{equation}
where $ \Phi_{m}^{\pm}=\varphi_{m}^{\pm}\left(  z_{0}\right)$.

The cross-mode coherence is given by%
\[
\left\langle a_{m}a_{n}^{\ast}\right\rangle =\Phi_{m}^{+}\Phi_{n}^{-}%
e^{-\frac{k^{2}}{2}\left\langle \left(  X_{m}^{+}-X_{n}^{+}\right)
^{2}\right\rangle }+\Phi_{m}^{+}\Phi_{n}^{+}e^{-\frac{k^{2}}{2}\left\langle
\left(  X_{m}^{+}-X_{n}^{-}\right)  ^{2}\right\rangle }%
\]%
\begin{equation}
+\Phi_{m}^{-}\Phi_{n}^{-}e^{-\frac{k^{2}}{2}\left\langle \left(  X_{m}%
^{-}-X_{n}^{+}\right)  ^{2}\right\rangle }+\Phi_{m}^{-}\Phi_{n}^{+}%
e^{-\frac{k^{2}}{2}\left\langle \left(  X_{m}^{-}-X_{n}^{-}\right)
^{2}\right\rangle }. \label{amn-mean}%
\end{equation}
In the particular case $m=n$, Eq. (\ref{amn-mean}) gives an expression for the
mean mode intensity%
\begin{equation}
\left\langle \left\vert a_{m}\right\vert ^{2}\right\rangle =2Q_{m}^{2}\left[
1+e^{-\frac{k^{2}}{2}\left\langle \left(  X_{m}^{+}-X_{m}^{-}\right)
^{2}\right\rangle }\sin(2kG_{m})\right]  , \label{am2-mean}%
\end{equation}
where%
\[
Q_{m}=q_{m}\left(  z_{0}\right)  =\frac{1}{\left[  D_{m}\tan\theta_{m}\left(
z_{0}\right)  \right]  ^{1/2}},
\]%
\[
G_{m}=g_{m}\left(  z_{0}\right)  =\frac{2\pi}{D_{m}}\int_{z_{\min}}^{z_{0}%
}\frac{dz}{\tan\left(  \theta_{m}\left(  z\right)  \right)  }.
\]
An expression for the mean squared intensity (the fourth moment of mode amplitude)
is
\[
\left\langle \left\vert a_{m}\right\vert ^{4}\right\rangle =Q_{m}^{4}\left[
6+8\sin\left(  2kG_{m}\right)  e^{-\frac{k^{2}}{2}\left\langle \left(
X_{m}^{+}-X_{m}^{-}\right)  ^{2}\right\rangle }\right.
\]%
\begin{equation}
-\left.  2\cos\left(  4kG_{m}\right)  e^{-2k^{2}\left\langle \left(  X_{m}%
^{+}-X_{m}^{-}\right)  ^{2}\right\rangle }\right]  . \label{am4-mean}%
\end{equation}

It is clear that similar formulas are readily derived for the joint
statistical moments of mode amplitudes at different frequencies.

In the scope of our ray-based approach, all the moments of mode amplitudes are
expressed through $\left\langle X_{m}^{\pm}X_{n}^{\mp}\right\rangle $ and
$\left\langle X_{m}^{\pm}X_{n}^{\pm}\right\rangle $. Evaluation of these
quantities is simplified under the assumption that the horizontal correlation
scale of sound speed fluctuations, $\Delta_{r}$, is substantially less than
the cycle length $D_{m}$. Then, at ranges $r\gg\Delta_{r}$ the mode rays cross uncorrelated inhomogeneities, the quantities $X_{m}^{+}$ and
$X_{m}^{-}$ become statistically independent, and $\left\langle X_{m}^{\pm
}X_{n}^{\mp}\right\rangle =0$.

Explicit expression for the dispersions of $X_{m}^{\pm}$ is given by
\[
\left\langle \left(  X_{m}^{\pm}\right)  ^{2}\right\rangle =\int_{0}^{r}%
\int_{0}^{r}\frac{dr^{\prime}dr^{\prime\prime}}{g_{m}\left(  r^{\prime
}\right)  g_{m}\left(  r^{\prime\prime}\right)  }%
\]

\begin{equation}
\times K\left(  r^{\prime}-r^{\prime\prime},z_{m}^{\pm}(r^{\prime})-z_{m}%
^{\pm}(r^{\prime\prime}),\frac{1}{2}\left(  z_{m}^{\pm}\left(  r^{\prime
}\right)  +z_{m}^{\pm}\left(  r^{\prime\prime}\right)  \right)  \right)  ,
\label{Xm2}%
\end{equation}
where $g_{m}\left(  r\right)  =\bar{c}\left(  z_{m}^{\pm}\left(  r\right)
\right)  \cos\left(  \chi_{m}^{\pm}\left(  r\right)  \right)  $. Let us change
the variables of integration from $\left(  r^{\prime},r^{\prime\prime}\right)
$ to $\left(  r^{\prime},\xi\right)  $, where $\xi=r^{\prime}-r^{\prime\prime}
$. The main contribution to the integral comes from the interval $\left\vert
\xi\right\vert <\Delta_{\zeta}$. Since $\Delta_{\zeta}$ is small compared to
$D_{m}$, we can use the approximations $z_{m}^{\pm}(r^{\prime})-z_{m}^{\pm
}(r^{\prime\prime})\simeq\xi\tan\chi_{m}^{\pm}\left(  r^{\prime}\right)  $ and
$\left(  z_{m}^{\pm}\left(  r^{\prime}\right)  +z_{m}^{\pm}\left(
r^{\prime\prime}\right)  \right)  /2\simeq z_{m}^{\pm}\left(  r^{\prime
}\right)  $. At ranges $r\gg\Delta_{\zeta}$ we can formally extend the limits
of integration over $\xi$ to infinity. Then Eq. (\ref{Xm2}) translates to
\begin{equation}
\left\langle \left(  X_{m}^{\pm}\right)  ^{2}\right\rangle =\int_{0}^{r}%
\frac{dr^{\prime}}{\bar{c}^{4}\left(  z_{m}^{\pm}\left(  r^{\prime}\right)
\right)  \cos^{2}\chi_{m}^{\pm}\left(  r^{\prime}\right)  }\int_{-\infty
}^{\infty}d\xi~K\left(  \xi,\xi\tan\chi_{m}^{\pm}\left(  r^{\prime}\right)
,z_{m}^{\pm}(r^{\prime})\right)  . \label{Xmp2}%
\end{equation}
Since the trajectories $z_{m}^{+}\left(  r\right)  $ and $z_{m}^{-}\left(
r\right)  $ differ only by a shift along the $r$-axis, the contribution to the
integral over $r^{\prime}$ from any interval of length $D_{m}$ is the same for
both mode rays. It is easy to see that for an arbitrary function $F(z_{m}%
^{\pm}(r^{\prime}),\left\vert \chi_{m}^{\pm}\left(  r^{\prime}\right)
\right\vert )$ we have the relation%
\[
\frac{1}{D_{m}}\int_{r_{1}}^{r_{1}+D_{m}}dr^{\prime}~F(z_{m}^{\pm}(r^{\prime
}),\left\vert \chi_{m}^{\pm}\left(  r^{\prime}\right)  \right\vert )=\frac
{2}{D_{m}}\int_{z_{\min}}^{z_{\max}}\frac{dz}{\tan\theta_{m}\left(  z\right)
}F\left(  z,\theta_{m}\left(  z\right)  \right)  .
\]
At ranges $r=ND_{m}$ with $N$ being an integer%
\begin{equation}
\left\langle \left(  X_{m}^{+}\right)  ^{2}\right\rangle =\left\langle \left(
X_{m}^{-}\right)  ^{2}\right\rangle =\left\langle X_{m}^{2}\right\rangle ,
\label{X2}%
\end{equation}
where%

\begin{equation}
\left\langle X_{m}^{2}\right\rangle =r\frac{2}{D_{m}}\frac{c_{0}^{2}k^{2}%
}{k_{m}^{2}}\int_{z_{\min}}^{z_{\max}}\frac{dz}{\bar{c}^{6}\left(  z\right)
\tan\theta_{m}\left(  z\right)  }\int_{-\infty}^{\infty}d\xi~K\left(  \xi
,\xi\tan\theta_{m}\left(  z\right)  ,z\right)  . \label{Xm2-smo}%
\end{equation}
In deriving this formula we have taken into account that due to Snell's law
$\cos\chi_{m}^{\pm}=(k_{m}/k)(\bar{c}(z_{m}^{\pm})/c_{0})$. At an arbitrary
range $r$, not necessary multiple of $D_{m}$, Eq. (\ref{Xm2-smo}) gives a
smoothed estimate of $\left\langle \left(  X_{m}^{+}\right)  ^{2}\right\rangle
$ and $\left\langle \left(  X_{m}^{-}\right)  ^{2}\right\rangle $.

Let us slightly simplify formulas (\ref{am-mean}) -- (\ref{am2-mean})
neglecting the differences between $\left\langle \left(  X_{m}^{+}\right)
^{2}\right\rangle $ and $\left\langle \left(  X_{m}^{-}\right)  ^{2}%
\right\rangle $ and between $\left\langle \left(  X_{m}^{+}-X_{n}^{+}\right)
^{2}\right\rangle $ and $\left\langle \left(  X_{m}^{-}-X_{n}^{-}\right)
^{2}\right\rangle $. Assuming that $\left\langle X_{m}^{\pm}X_{m}^{\mp
}\right\rangle =0$ and replacing $\left\langle \left(  X_{m}^{\pm}\right)
^{2}\right\rangle $ by $\left\langle \left(  X_{m}\right)  ^{2}\right\rangle $
and $\left\langle \left(  X_{m}^{\pm}-X_{n}^{\pm}\right)  ^{2}\right\rangle $
by $\left\langle \left(  X_{m}-X_{n}\right)  ^{2}\right\rangle $, we find%
\begin{equation}
\left\langle a_{m}\right\rangle =\varphi_{m}(z_{0})e^{-\frac{k^{2}}%
{2}\left\langle X_{m}^{2}\right\rangle }, \label{am}%
\end{equation}%
\[
\left\langle a_{m}a_{n}^{\ast}\right\rangle =2Q_{m}Q_{n}\left[  e^{-\frac
{k^{2}}{2}\left\langle \left(  X_{m}-X_{n}\right)  ^{2}\right\rangle }%
\cos\left(  k\left(  G_{m}-G_{n}\right)  \right)  \right.
\]%
\begin{equation}
\left.  +e^{-\frac{k^{2}}{2}\left(  \left\langle X_{m}^{2}\right\rangle
+\left\langle X_{n}^{2}\right\rangle \right)  }\sin(k\left(  G_{m}%
+G_{n}\right)  )\right]  , \label{amn}%
\end{equation}%
\begin{equation}
\left\langle \left\vert a_{m}\right\vert ^{2}\right\rangle =2Q_{m}^{2}\left[
1+e^{-k^{2}\left\langle X_{m}^{2}\right\rangle }\sin(2kG_{m})\right]  ,
\label{am2}%
\end{equation}
and%
\[
\left\langle \left\vert a_{m}\right\vert ^{4}\right\rangle =Q_{m}^{4}\left[
6+8\sin\left(  2kG_{m}\right)  e^{-k^{2}\left\langle X_{m}^{2}\right\rangle
}\right.
\]%
\begin{equation}
-\left.  2\cos\left(  4kG_{m}\right)  e^{-4k^{2}\left\langle X_{m}%
^{2}\right\rangle }\right]  . \label{am4}%
\end{equation}
The value of $\left\langle X_{m}^{2}\right\rangle $ is given by Eq.
(\ref{Xm2-smo}). We do not present an explicit expression for $\left\langle
\left(  X_{m}-X_{n}\right)  ^{2}\right\rangle $. When using Eq. (\ref{amn}),
$\exp\left[  -\frac{k^{2}}{2}\left\langle \left(  X_{m}-X_{n}\right)
^{2}\right\rangle \right]  $ should be replaced by any of two close functions
$\exp\left[  -\frac{k^{2}}{2}\left\langle \left(  X_{m}^{\pm}-X_{n}^{\pm
}\right)  ^{2}\right\rangle \right]  $.

\section{Ray-based approach and master equations \label{sec:master}}

The equations for statistical moments $\left\langle a_{m}a_{n}^{\ast
}\right\rangle $ are derived in the Markov approximation
proceeding from the mode coupling equation \cite{DT1,Creamer1996,Colosi2009}%

\begin{equation}
\frac{da_{m}}{dr}=i\sum_{n=1}^{M}\rho_{mn}a_{n}e^{ik_{nm}r}, \label{amplitude}%
\end{equation}
where $k_{nm}=k_{n}-k_{m}$,%
\begin{equation}
\rho_{mn}(r)=\frac{k^{2}}{\sqrt{k_{m}k_{n}}}\int dz~\varphi_{m}(z)\mu\left(
r,z\right)  \varphi_{n}(z), \label{rho-def}%
\end{equation}%
\begin{equation}
\mu\left(  r,z\right)  =-\frac{c_{0}^{2}}{\bar{c}^{3}\left(  z\right)  }\delta
c\left(  r,z\right)  . \label{mu-def}%
\end{equation}
Note that our main formula (\ref{am-ray}) is an approximate solution of Eq.
(\ref{amplitude}) \cite{V87}.

In Refs. \cite{Colosi2009,Colosi2012,Colosi2014}, it was shown that the numerical
solution of the complete system of equations for all the moments $\left\langle
a_{m}a_{n}^{\ast}\right\rangle $ give practically the same result as the
evaluation of these moments in the Monte Carlo simulation based on numerical
solving the mode coupling equation (\ref{amplitude}) for different
realizations of random perturbation $\delta c(r,z)$. This was the expected
result. An unexpected result was that even though the cross-mode coherences
were not small, the master equations properly predicted the smoothed mode
intensities. In this section we will show that this result follows from our
ray-based estimates of statistical moments. Namely, it will be shown that Eq.
(\ref{am2}) obtained by smoothing the range-dependent parameters of Eq.
(\ref{am2-mean}), in the limit of high frequency gives a solution to the
master equations.

In the notation of Refs. \cite{Colosi2009,Colosi2012,Colosi2014} the master
equations have the form
\begin{equation}
\frac{d\left\langle \left\vert a_{m}\right\vert ^{2}\right\rangle }{dr}%
=2\sum_{n=0}^{M}\operatorname{Re}\left(  I_{mn,nm}\right)  \left(
\left\langle \left\vert a_{n}\right\vert ^{2}\right\rangle -\left\langle
\left\vert a_{m}\right\vert ^{2}\right\rangle \right)  , \label{coupling}%
\end{equation}
where $I_{mn,nm}$ are the elements of the scattering matrix defined by the
relations%
\begin{equation}
I_{mn,qp}=\int_{0}^{\infty}d\xi~\Delta_{mn,qp}\left(  \xi\right)
e^{ik_{pq}\xi} \label{Imn-def}%
\end{equation}
and%

\begin{equation}
\Delta_{mn,pq}\left(  r-r^{\prime}\right)  =\left\langle \rho_{mn}(r)\rho
_{pq}\left(  r^{\prime}\right)  \right\rangle . \label{D-def}%
\end{equation}

According to Eqs. (\ref{rho-def}), (\ref{mu-def}), and (\ref{D-def}),%
\[
\Delta_{mn,nm}\left(  \xi\right)  =\frac{k^{4}c_{0}^{4}}{k_{n}k_{m}}\int
dzdz^{\prime}~\varphi_{n}(z)\varphi_{n}(z^{\prime})
\]

\begin{equation}
\times\frac{K\left(  \xi,z-z^{\prime},\left(  z+z^{\prime}\right)  /2\right)
}{\bar{c}^{3}\left(  z\right)  \bar{c}^{3}\left(  z^{\prime}\right)  }%
\varphi_{m}(z)\varphi_{m}(z^{\prime}). \label{Imn-ray}%
\end{equation}

At high frequencies, where the wavelength is small compared to the spatial
scales of perturbation $\delta c$, we deal with the small-angle forward
scattering of sound waves, and each mode couples mainly into modes with close
numbers. This means that the main contribution to the sum in the right hand
side of Eq. (\ref{amplitude}) comes from terms with $n$ close to $m$. Using
Eq. (\ref{phi-sum}), we present the product of two eigenfunctions with close
$m$ and $n$ in the form%
\begin{equation}
\varphi_{n}\left(  z\right)  \varphi_{m}\left(  z\right)  \simeq q_{m}%
^{2}\left(  z\right)  \left\{  e^{ik\left[  g_{n}\left(  z\right)
-g_{m}\left(  z\right)  \right]  }+c.c.\right\}  , \label{phi-mn}%
\end{equation}
where $c.c.$ denotes the complex conjugate of the preceding term.
In the right hand side of Eq. (\ref{phi-mn}) we have omitted
rapidly oscillating terms whose contributions to the integral in
Eq. (\ref{Imn-ray}) are negligible. The integrand on the right of
Eq. (\ref{Imn-ray}) is non-negligible only for $\left\vert
z-z^{\prime}\right\vert =O\left(  \Delta_{\zeta}\right)  $. We
assume that the vertical scale of perturbation $\Delta_{\zeta}$ is
small compared to the depth interval between turning points of the
$m$-th mode. Then the Brillouin waves within the depth interval of
width $\Delta_{\zeta}$ can be approximated by plane waves.
Substitute Eq. (\ref{phi-mn}) in Eq. (\ref{Imn-ray}) and drop the
rapidly oscillating terms. Using Eqs. (\ref{quantization-diff})
and (\ref{dGm-dm}), the phases of the remaining terms can be
represented as
\[
k\left[  g_{n}\left(  z\right)  -g_{m}\left(  z\right)  -g_{n}\left(
z^{\prime}\right)  +g_{m}\left(  z^{\prime}\right)  \right]  \simeq
k\frac{\partial^{2}g_m\left(z\right)  }{\partial z\partial m}\left(
n-m\right)  \left(  z^{\prime}-z\right)
\]%
\[
=\frac{2\pi\left(  n-m\right)  \left(  z-z^{\prime}\right)  }{D_{m}}\cot
\theta_{m}\left(  z\right)  .
\]
In the resulting expression, we approximately replace $n$ and $z^{\prime}$ in
the pre-exponential factors by $m$ and $z$, respectively. This yields%
\[
\Delta_{nm,mn}\left(  \xi\right)  =\frac{k^{4}c_{0}^{4}}{k_{m}^{2}}%
\int_{z_{\min}}^{z_{\max}}dz\int_{-\infty}^{\infty}d\zeta
\]

\begin{equation}
\times\frac{K\left(  \xi,\zeta,z\right)  }{\bar{c}^{6}\left(  z\right)
D_{m}^{2}\tan^{2}\theta_{m}\left(  z\right)  }~\left[  e^{\frac{2\pi i\left(
n-m\right)  }{D_{m}}\zeta\cot\theta_{m}\left(  z\right)  }+c.c.\right]  .
\label{Dmn}%
\end{equation}
Let us plug Eq. (\ref{Dmn}) into Eq. (\ref{Imn-def}) and use the relation
$k_{nm}=2\pi\left(  m-n\right)  /D_{m}$ which follows from Eq.
(\ref{quantization-diff}). At high frequencies, the number of propagating mode
becomes very large and the sum $\sum_{n}\ldots$ in Eq. (\ref{coupling}) can be
approximately replaced by the integral $\int dn\ldots$. For $m$ satisfying the
condition $1\ll m\ll M$, we will use the approximate relation%
\[
\int_{0}^{M}dn~\exp\left[  \frac{2\pi i\left(  m-n\right)  }{D_{m}}\left(
\xi\pm\zeta\cot\theta_{m}\right)  \right]  =D_{m}\delta\left(  \xi\pm\zeta
\cot\theta_{m}\right)  .
\]
Then%
\[
\sum_{n=0}^{M}\operatorname{Re}I_{mn,nm}=\int_{0}^{\infty}d\xi\int_{0}%
^{M}dn~\Delta_{mn,nm}\left(  \xi\right)  \cos\left[  \frac{2\pi i\left(
n-m\right)  }{D_{m}}\xi\right]
\]%
\begin{equation}
=\frac{k^{4}c_{0}^{4}}{k_{m}^{2}D_{m}}\int\frac{dz}{\bar{c}^{6}\left(
z\right)  \tan\theta_{m}\left(  z\right)  }\int_{0}^{\infty}d\xi~\left[
K\left(  \xi,\xi\tan\theta_{m},z\right)  +K\left(  \xi,-\xi\tan\theta
_{m},z\right)  \right]  . \label{sim_I}%
\end{equation}
From the comparison of this expression with Eq. (\ref{Xm2-smo}) (taking into
account Eq. (\ref{K-sym})) we find%
\begin{equation}
2\sum_{n=0}^{M}\operatorname{Re}\left(  I_{mn,nm}\right)  =k^{2}%
\frac{d\left\langle X_{m}^{2}\right\rangle }{dr}. \label{dXm2-dr}%
\end{equation}

Plugging Eq. (\ref{am2}) into the left-hand side (l.h.s.) and right-hand side
(r.h.s.) of Eq. (\ref{coupling}) yields:%

\begin{equation}
\text{l.h.s.}=2k^{2}Q_{m}^{2}\sin\left(  2kG_{m}\right)  e^{-k^{2}\left\langle
X_{m}^{2}\right\rangle }\frac{d\left\langle X_{m}^{2}\right\rangle }{dr},
\label{lhs}%
\end{equation}
and%
\begin{equation}
\text{r.h.s.}=A+B+C, \label{rhs}%
\end{equation}
where%
\begin{equation}
A=4Q_{m}^{2}e^{-k^{2}\left\langle X_{m}^{2}\right\rangle }\sin(2kG_{m}%
)\sum_{n=0}^{M}\operatorname{Re}\left(  I_{mn,nm}\right)  , \label{A}%
\end{equation}%
\begin{equation}
B=4\sum_{n=0}^{M}\operatorname{Re}\left(  I_{mn,nm}\right)  \left(  Q_{n}%
^{2}-Q_{m}^{2}\right)  , \label{B}%
\end{equation}%
\begin{equation}
C=-4\sum_{n=0}^{M}\operatorname{Re}\left(  I_{mn,nm}\right)  Q_{n}^{2}%
\sin(2kG_{n})e^{-k^{2}\left\langle X_{n}^{2}\right\rangle }. \label{C}%
\end{equation}
According to Eq. (\ref{dXm2-dr}), l.h.s. $=A$. This means that l.h.s. = r.h.s.
if term $A$ dominates in sum (\ref{rhs}). $\ $

According to Eq. (\ref{am2}), the mean intensity $\left\langle \left\vert
a_{m}\right\vert ^{2}\right\rangle $ varies at ranges where $k^{2}\left\langle
X_{m}^{2}\right\rangle =O\left(  1\right)  $. It can be shown that at these
ranges and at sufficiently large $k$ the term $A$ dominates in the sum
(\ref{rhs}). The smallness of term $B$ is caused by the fact that $Q_{m}^{2}$ is a smooth
function of the mode number $m$. Analysis of Eqs. (\ref{Imn-def}) and
(\ref{Dmn}) shows that the values of $\Delta_{mn,nm}$ and $I_{mn,nm}$
descrease with increasing $\left\vert n-m\right\vert $ and the main
contributions to sums (\ref{A}) and (\ref{B}) come from terms with $n$
belonging to some interval $\left\vert n-m\right\vert <\Delta m$. Consider
Brillouin waves with grazing angles close to some fixed value. It is easy to
show that the numbers $m$ of corresponding modes grow with frequency but the
values of $Q_{m}$ and $\Delta m$ for these modes will be approximately
constant. It means that $\left\vert Q_{m+\Delta m}^{2}-Q_{m}^{2}\right\vert
/Q_{m}^{2}=O(1/k)$ and therefore in the high frequency limit the ratio $B/A$
tends to zero.

The smallness of $C$ compared to $A$ is caused by the presence in Eq.
(\ref{C}) a rapidly oscillating factor $\sin(2kG_{n})$. This can be shown by
transforming sum (\ref{C}) in the same manner as it has been done for sum
$\sum_{n}\operatorname{Re}I_{mn,nm}$.

\section{ Numerical example \label{sec:numeric}}

In this Section we present results of numerical simulation demonstrating the
applicability of Eq. (\ref{am-ray}) and estimates of statistical moments
obtained using this formula. As in Ref. \cite{Colosi2009} we consider a
deep-water waveguide with the canonical sound speed profile and perturbation
$\delta c\left(  r,z\right)  $ induced by random internal waves.

\subsection{Environmental model and numerical simulation \label{sub:model}}

In numerical simulations presented below we use an environmental model with an
unperturbed sound speed profile representing the canonical (or Munk) profile
\cite{BL2003,FDMWZ79}%
\begin{equation}
\bar{c}(z)=c_{r}\left[  1+\varepsilon\left(  e^{\eta}-\eta-1\right)  \right]
,\;\;\eta=2(z_{a}-z)/B \label{Munk}%
\end{equation}
with parameters $c_{r}=1.5$ km/s, $\varepsilon=0.0057$, $B=1$ km, and
$z_{a}=1$ km. This $\bar{c}(z)$ is shown in the left panel of Fig.
1. The bottom was set at a depth of 5 km.

It is assumed that the weak perturbation $\delta c(r,z)$ is caused by random
internal waves with statistics determined by the empirical Garrett-Munk
spectrum \cite{FDMWZ79}. To generate realizations of a random field $\delta
c(r,z)$ we apply a numerical technique developed by J. Colosi and M. Brown
\cite{CB98}. In their model the perturbation has the form%
\begin{equation}
\delta c(r,z)=c_{r}\frac{\mu}{g}N^{2}\zeta(r,z), \label{dc-appendix}%
\end{equation}
where $g=9.8$ m/s$^{2}$ is the gravitational acceleration, $\mu=24.5$ is a
dimensionless constant, $N(z)=N_{0}\exp(z/L)$ is a buoyancy frequency profile,
$N_{0}=2\pi/(12$ min$)=0.0087$ 1/s is a buoyancy frequency near the surface,
$L=1$ km. The random function $\zeta(r,z)$ presents internal-wave-induced
vertical displacements of a fluid parcel. Its realizations have been computed
using Eq. (19) from Ref. \cite{CB98}. We consider an internal wave field
formed by $30$ normal modes and assume its horizontal isotropy. Components of
wave number vectors in the horizontal plane belong to the interval from
$2\pi/100$ km$^{-1}$ to $2\pi/2$ km$^{-1}$. An rms amplitude of the
perturbation scales in depth like $\exp(3z/2L)$ and its surface-extrapolated
value in our model is about $0.5$ m/s.

All the calculations were carried out at a carrier frequency of
100 Hz. The point source exciting the wave field was set at the
sound channel axis $z=z_{a}$. The complex amplitude of the wave
field was computed using the method of wide angle parabolic
equation for 360 realizations of random perturbation $\delta
c\left(  r,z\right)  $. Parabolic equation was solved by applying
the Crank--Nicolson finite-difference scheme \cite{JKPS2011}. For
each realization of $\delta c(r,z)$, the complex amplitude of the
sound field was computed up to 500 km. Starting field at $r=0$ was
generated using the modal starter \cite{JKPS2011}. Values of
$a_{m}$ at 501 range points uniformly sampling the interval from 0
to 500 km were found by projecting the computed sound field onto
eigenfunctions $\varphi_{m}\left(  z\right)  $. Functions
$a_{m}\left(  r\right)  $ obtained this way we call the pe-based
estimates of mode amplitudes.

Our attention was restricted to amplitudes of the first 66 modes which
describe sound waves propagating at grazing angle $\left\vert \chi\right\vert
<11.6^{\circ}$. Turning points of these modes are located within the water
bulk and far enough from the boundaries for the applicability of quantization
rule (\ref{quantization}). Starting intensities $\left\vert a_{m}\left(
0\right)  \right\vert ^{2}$ of some of these modes are shown by circles in
Fig. 2. Numbers of modes whose statistical moments will be shown on the plots
presented below, are indicated next to the corresponding circles.

\begin{figure}[ptb]
\begin{center}
\includegraphics[
width=11cm
]{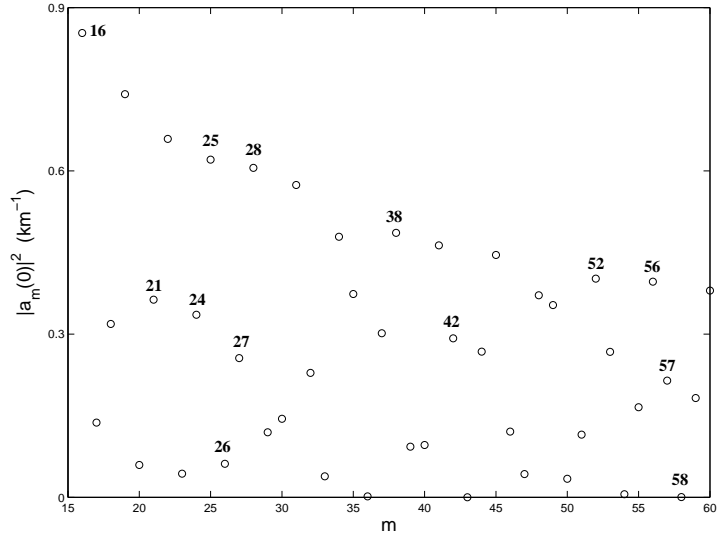}
\end{center}
\par
\caption{Starting intensities $\left\langle \left\vert a_{m}\left(  0\right)
\right\vert ^{2}\right\rangle $ of normal modes with $m=16,...,60$ at a
carrier frequency of 100 Hz excited by a point source set at a depth of 1 km.
}%
\label{fig_Im0}%
\end{figure}

Functions $X_{m}^{+}(r)$ and $X_{m}^{-}(r)$ determined by Eq. (\ref{Xmp1})
were computed for the same 360 realizations of $\delta c\left(  r,z\right)  $
at the same 501 range points. Then, substituting these functions in Eq.
(\ref{am-ray}) we obtained the ray-based estimates of $a_{m}\left(  r\right)
$.

Thus, in each of 360 realizations of $\delta c(r,z)$ for each of the first 66
modes we computed a pe-based and ray-based estimate of mode amplitude
$a_{m}(r)$. For most modes these two estimates of $a_{m}\left(  r\right)  $
are in reasonable agreement. Figure 3 present typical
examples of $\left\vert a_{m}\left(  r\right)  \right\vert $ computed these
two ways for the same realization of perturbation.

\begin{figure}[ptb]
\begin{center}
\includegraphics[
width=12.2cm
]{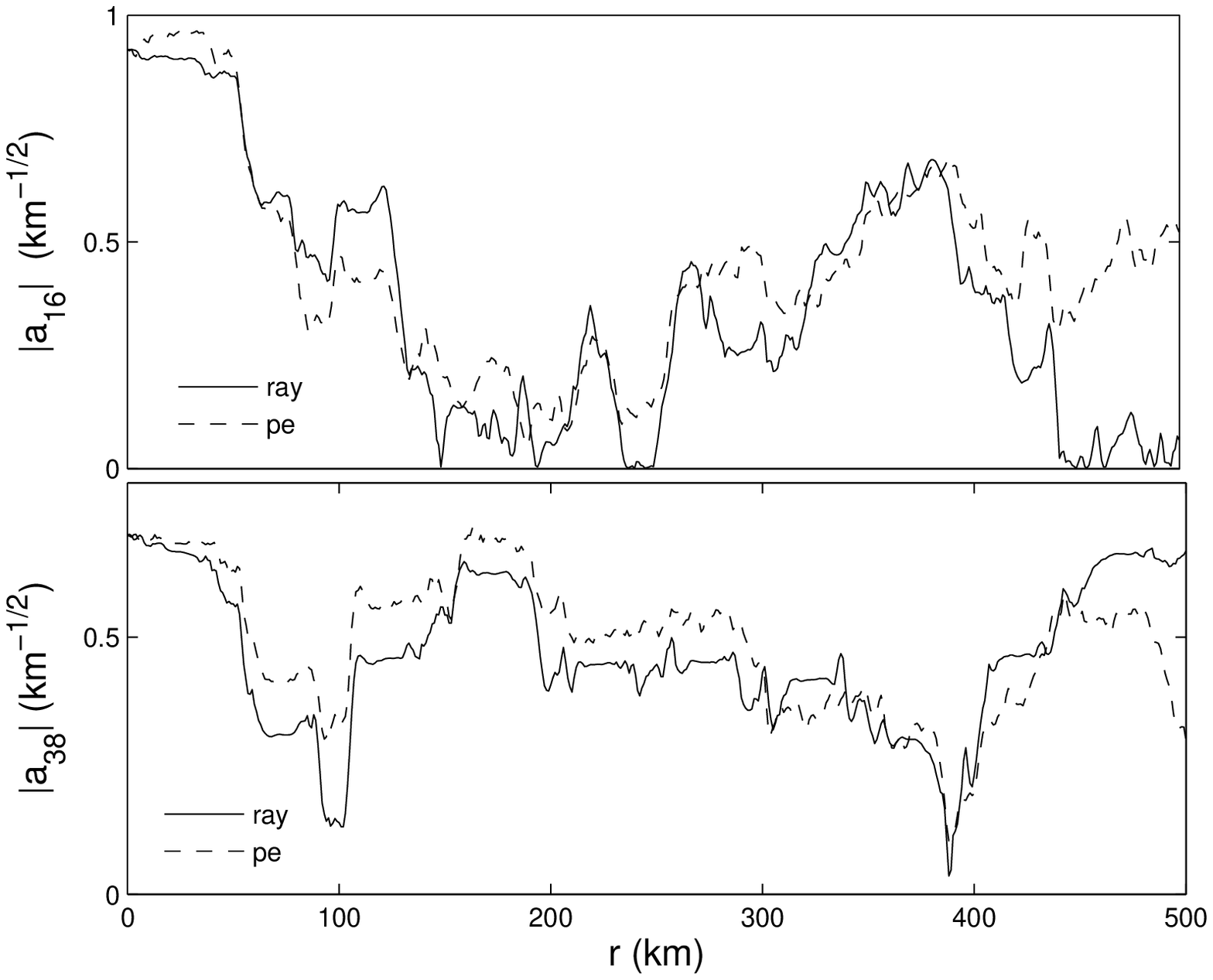}
\end{center}
\caption{Amplitudes $\left\vert a_{m}\right\vert $ of the 16-th \ (upper
panel) and 38-th (lower panel) modes as functions of range in a single
realization of a fluctuating waveguide. The ray-based estimate predicted by
Eq. (\ref{am-ray}) (solid line) is compared to result obtained by modal
decomposition of the sound field computed by the method of wide-angle
parabolic equation (dashed line).}%
\label{fig_single_realiz}%
\end{figure}

\subsection{Statistical moments of $a_{m}$ \label{sub:stat}}

In the remaing part of this paper we will compare the pe-based and ray-based
estimates of statistical moments calculated using the Monte Carlo method. In
what follows, the angular brackets $\left\langle \ldots\right\rangle $ denote
the averaging over the 360 realizations of perturbation $\delta c$.

The ray-based estimates of statistical moments can be obtained in two ways.
First, we can substitute functions $X_{m}^{\pm}(r)$ computed for different
realizations of $\delta c$ into formula (\ref{am-ray}) and average a product
of mode amplitudes over all the realizations. Second, we can find the second
moments of $X_{m}^{\pm}(r)$ by averaging over the realizations and substitute
these moments in Eqs. (\ref{am-mean})--(\ref{am4-mean}). Both methods give
close results. Therefore below we present only the estimates by the first method.

Figure 4 shows the mean values (coherent components) of
complex mode amplitudes $\left\vert \left\langle a_{m}\right\rangle
\right\vert $ as functions of range $r$. Comparison with similar dependencies
for non-averaged mode amplitudes presented in Fig. 3
show that the averaging makes the pe-based and ray-based results more close.

\begin{figure}[ptb]
\begin{center}
\includegraphics[
width=12cm
]{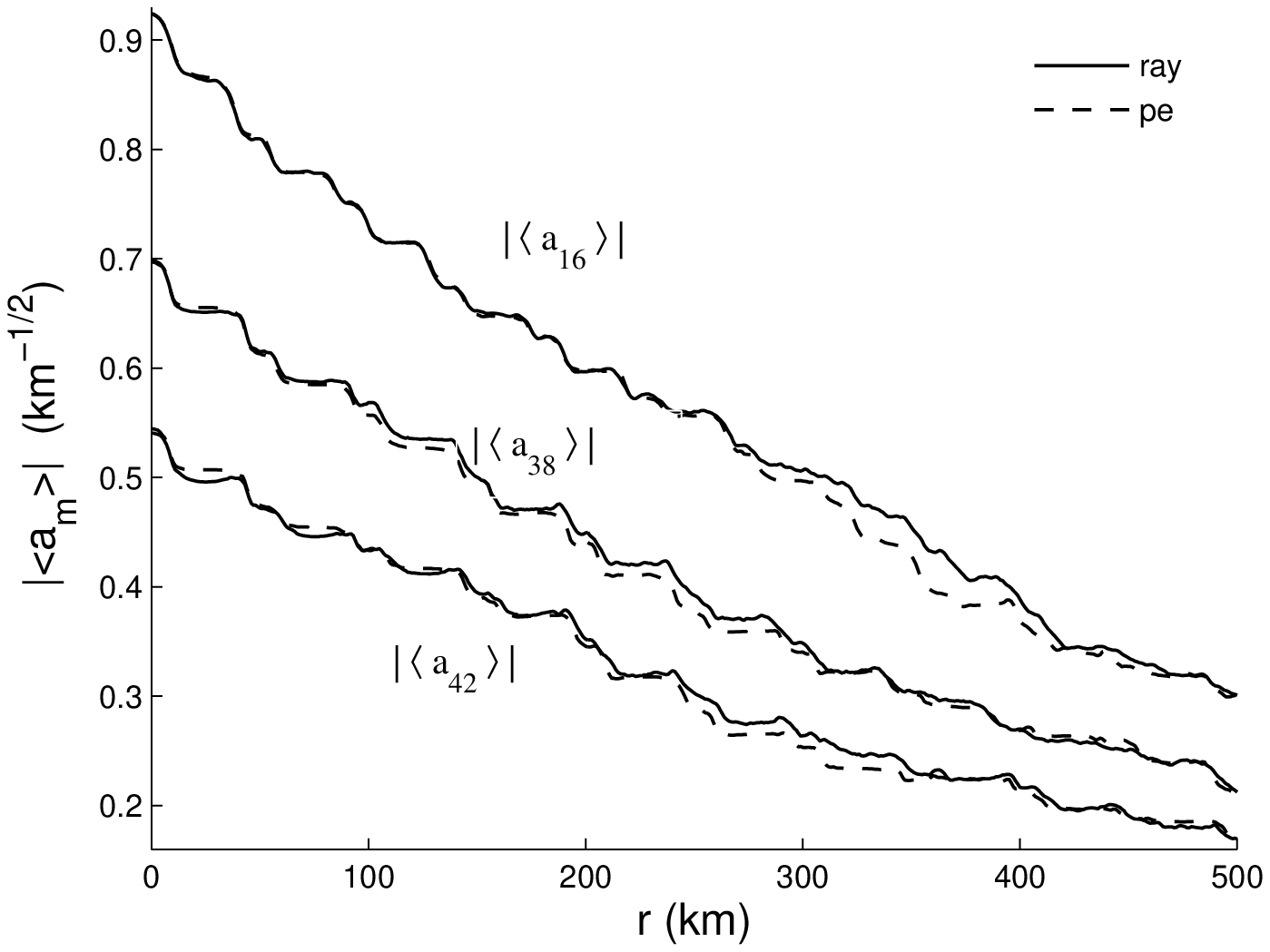}
\end{center}
\caption{Ray-based (solid lines) and pe-based (dashed lines) estimates of mean
mode amplutudes $\left\vert \left\langle a_{m}\right\rangle \right\vert $ for
$m$ = 16 (upper lines), 38 (middle lines), and 43 (lower lines).}%
\label{fig_coherent}%
\end{figure}

As is seen in Fig. 2, the starting mode intensity $\left\vert
a_{m}\left(  0\right)  \right\vert ^{2}$ is a rapidly oscillating function of
the mode number $m$. For $m>3$ the values of $\left\vert a_{m}\left(
0\right)  \right\vert ^{2}$ in our example are well approximated by the WKB
relation%
\begin{equation}
\left\vert a_{m}\left(  0\right)  \right\vert ^{2}=2Q_{m}^{2}\left[
1-\sin(2kG_{m})\right]  . \label{am0}%
\end{equation}
According to this formula, the oscillations are caused by term $\sin(2kG_{m}%
)$. In Eqs. (\ref{am2-mean}) and (\ref{am2}) the mode coupling
manifests itself in the appearance of a weight factor at
$\sin(2kG_{m})$ which monotonically decreases with range. It means
that mean intensities of modes with the numbers close to $m$
monotonically approaches to $2Q_{m}^{2}$. The equalization of mean
mode intensities is clearly seen in Fig. 5 where the range
dependencies of mean intensities for modes 56, 57, and 58 are
shown. Note that the pe-based (thick solid) and ray-based (thick
dash) simulations give results close to each other and to the
solution of master equations (\ref{coupling}) (thin solid). In accord with
results of Ref. \cite{Colosi2009,Colosi2012,Colosi2014} and our
result derived in Sec. \ref{sec:master}, the solution of Eqs.
(\ref{coupling}) gives only a smoothed range dependence of the
mean mode intensity $\left\langle \left\vert a_{m}\right\vert
^{2}\right\rangle $ and does not describe its small oscillations.

\begin{figure}[ptb]
\begin{center}
\includegraphics[
width=12cm
]{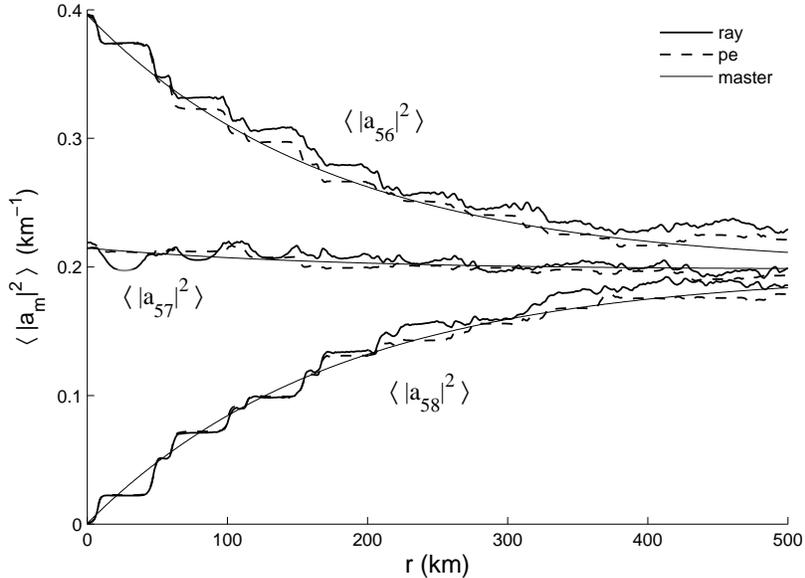}
\end{center}
\caption{Mean mode intensities $\left\langle \left\vert a_{m}\right\vert
^{2}\right\rangle $ as functions of range for $m$ = 56, 57, and 58. The
ray-based results, pe-based results, and solutions of the master equations are
shown by thick solid, thick dashed, and thin solid lines, respectively. }%
\label{fig_mode_intensity}%
\end{figure}

The presence of these oscillations was predicted in Ref. \cite{V89}. From the
viewpoint of our ray-based approach they are caused by the fact that the
strength of the sound speed fluctuations decreases with depth and therefore
the main contributions to integrals (\ref{Xmp-def}) come from inhomogeneities
located in the vicinities of upper turning points of the mode rays.
Near-step-like jumps of $X_{m}^{\pm}\left(  r\right)  $ occur at the mode
rays' upper turning points. The same is true for the random increment $X(r)$
of eikonal of any geometrical ray described by Eq. (\ref{X-geom}). This fact
is well known and it underlies the so-called apex approximation \cite{FDMWZ79}%
. In our example this effect is most pronounced for steep enough rays with
grazing angles at the sound channel axis exceeding 5$^{\circ}$. These rays
form modes with numbers $m\geq12$.

According to Eqs. (\ref{am-mean}) -- (\ref{am2-mean}) the jump-like variations
of $X_{m}^{\pm}\left(  r\right)  $ at mode rays' upper turning points cause
jump-like variations of statistical moments at the corresponding distances.
This phenomenon is illustrated in Fig. 6 where the range
dependencies of the mean amplitude (upper panel) and mean intensity (middle
panel) of the 52-nd mode are shown. Dashed vertical lines indicate ranges
corresponding to upper turning points of mode rays of the 52-nd mode depicted
in the lower panel.

\begin{figure}[ptb]
\begin{center}
\includegraphics[
width=12cm
]{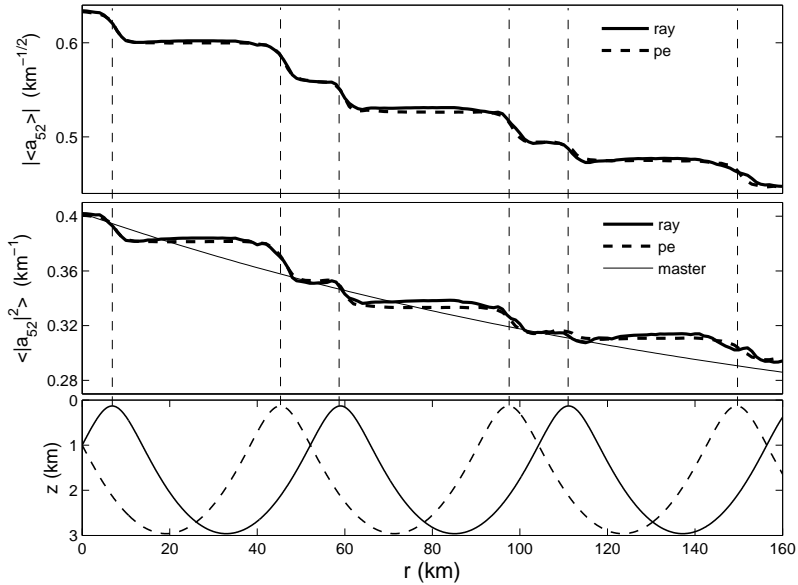}
\end{center}
\caption{Upper panel: mean amplitude of the 52-nd mode as a function of range.
Middle panel: mean intensity of the 52-nd mode as a function of range. Lower
panel: mode rays of the 52-nd modes. Thick solid and thick dashed lines in the
upper and middle panels show the ray-based and pe-based results. Thin solid
line in the middle panel presents the mean intensity $\left\langle \left\vert
a_{52}\right\vert ^{2}\right\rangle $ found by solving master equations
(\ref{coupling}).}%
\label{fig_apex}%
\end{figure}

Formulas (\ref{am})--(\ref{am4}) are derived using the
approximation of $\left\langle \left(  X_{m}^{\pm}\left(  r\right)
\right)  ^{2}\right\rangle $ by the smooth function $\left\langle
X_{m}^{2}\left(  r\right) \right\rangle $ which has no jumps at
the upper turning points of mode rays. This explains why the
master equations (\ref{coupling}), whose soulutions (in the high
frequency approximation) are given by Eq. (\ref{am2}), do not
predict the small oscillations of mean mode intensities
$\left\langle \left\vert a_{m}\left(  r\right)  \right\vert
^{2}\right\rangle $.

Figure 7 presents the mean squared intensities of the same modes
as in Fig. 5. It is seen that the agreement between the ray-based
and pe-based estimates for the fourth moments of the mode
amplitudes is less good than for the second moments.

\begin{figure}[ptb]
\begin{center}
\includegraphics[
width=12cm
]{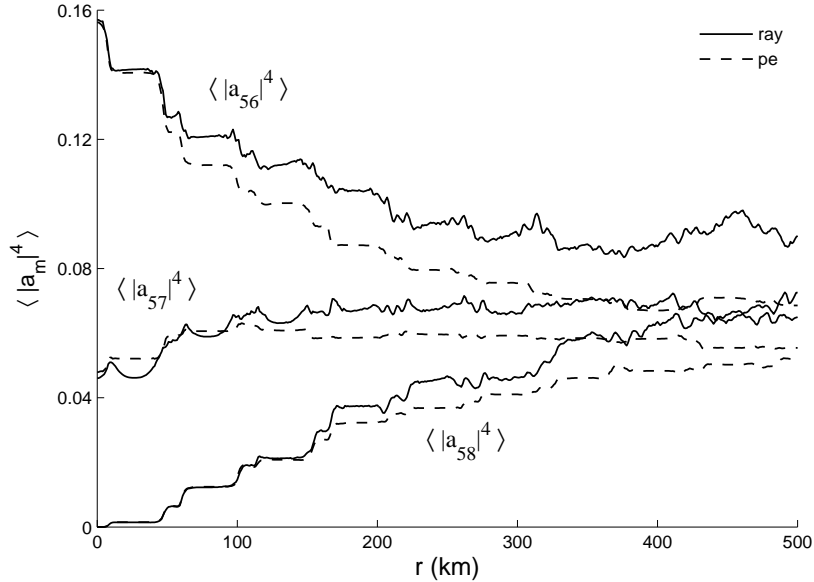}
\end{center}
\caption{Squared mode intensities for the same normal modes as in Fig. 5. The
ray-based and pe-based results are shown by solid and dashed lines,
respectively. }%
\label{fig_a4}%
\end{figure}

\subsection{ Cross-correlations of normal modes \label{sub:cross}}

According to Eqs. (\ref{amn-mean}), the decorrelation of modes $m$ and $n$ is
determined by functions%
\begin{equation}
Y_{m,n}^{\pm}(r)=\exp\left( - \frac{k^2}{2} \left\langle \left[  X_{m}^{\pm}(r)-X_{n}^{\pm
}\left(  r\right)  \right]  ^{2}\right\rangle \right)  . \label{Ymn}%
\end{equation}
We will call $Y_{m,n}^{\pm}$ the correlation functions of mode rays. Figure
8 presents the values of $Y_{mn}^{+}$ at 125 (upper
panel), 250 (middle panel), and 500 km (lower panel). Functions $Y_{mn}^{-}$
have close values (not shown). Let us assume that modes $m$ and $n$ are
correlated if $Y_{m,n}^{+}>0.6$. Then in Fig. 8, we
see that at 125 km a typical mode correlates with 20-40 neighboring modes, at
250 km the number of correlated modes reduces to 10-20, and at 500 km to 3-5.

\begin{figure}[ptb]
\begin{center}
\includegraphics[
width=12cm
]{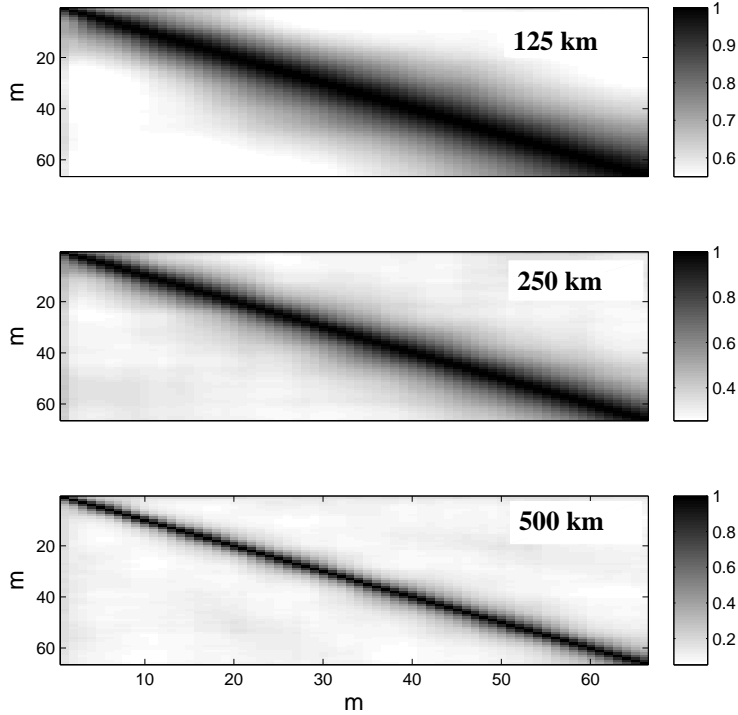}
\end{center}
\caption{Correlation functions of mode rays $Y_{mn}^{+}$ (Eq. (\ref{Ymn})) at
ranges 125 km (upper panel), 250 km (middle panel), 500 km (lower panel).}%
\label{fig_ray_correlation}%
\end{figure}

Correlation functions of mode rays $Y_{m,n}^{\pm}(r)$ monotonically decrease
with increasing $r$ and $\left\vert m-n\right\vert $. But, according to Eqs.
(\ref{amn-mean}) and (\ref{amn}), the dependencies of joint statistical
moments $\left\langle a_{m}a_{n}^{\ast}\right\rangle $ on $r$ and $\left\vert
m-n\right\vert $ can be more complicated. In the analysis of the cross-mode
coherence, as in Refs. \cite{Colosi2009,Colosi2012,Colosi2014}, we will
consider the normalized joint moments of mode amplitudes $\left\langle
b_{m}b_{n}^{\ast}\right\rangle $, where%

\[
b_{m}\left(  r\right)  =\frac{a_{m}\left(  r\right)  }{\left\langle \left\vert
a_{m}\left(  r\right)  \right\vert ^{2}\right\rangle ^{1/2}}.
\]
In the upper panel of Fig. 9 we present the range dependencies
of cross-mode correlations for modes with very different starting amplitudes
at $r=0$ (cf. Fig. 2). Joint moments $\left\vert \left\langle
b_{m}b_{n}^{\ast}\right\rangle \right\vert $ of such modes within some range
intervals may grow with range $r$ and increase with increasing $\left\vert
m-n\right\vert $. In our example we see that $\left\vert \left\langle
b_{25}b_{27}^{\ast}\right\rangle \right\vert $ grows in the interval from 100
to 400 km, and $\left\vert \left\langle b_{25}b_{28}^{\ast}\right\rangle
\right\vert $ exceeds $\left\vert \left\langle b_{25}b_{26}^{\ast
}\right\rangle \right\vert $ and $\left\vert \left\langle b_{25}b_{27}^{\ast
}\right\rangle \right\vert $.

\begin{figure}[ptb]
\begin{center}
\includegraphics[
width=12cm
]{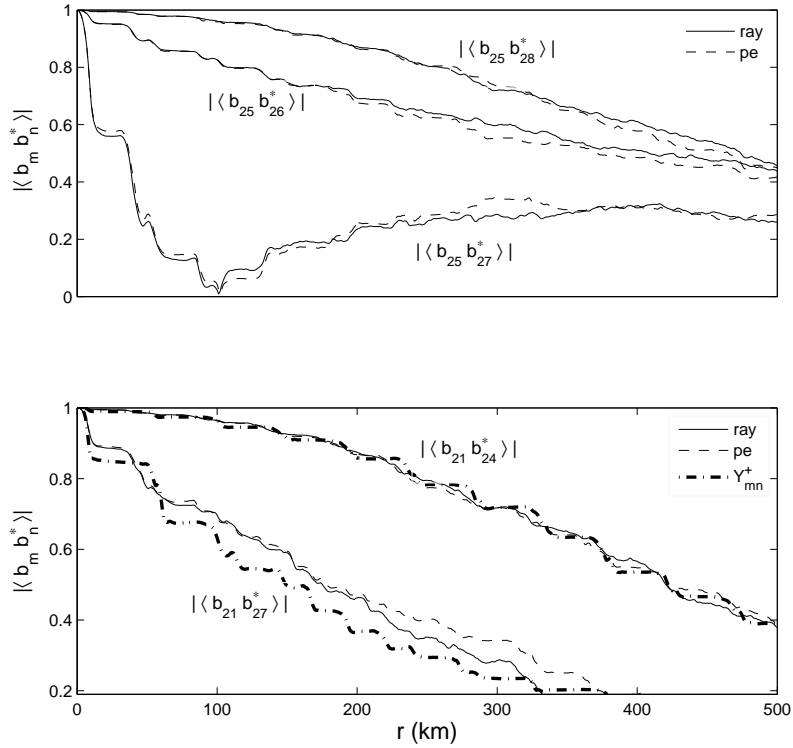}
\end{center}
\caption{Cross-mode coherences $\left\vert \left\langle b_{m}b_{n}^{\ast
}\right\rangle \right\vert $ as functions of range. Coherences for mode pairs
(25,26), (25,27), and (25,28) are shown in the upper panel; for mode pairs
(21,24) and (21,27) are shown in the lower panel. In both panels, the
ray-based and pe-based results are shown by thin solid and thin dash lines,
respectively. In the lower panel, the correlation functions of mode rays
$Y_{mn}^{+}$ (Eq. (\ref{Ymn})) are shown by thick dot-dashed lines.}%
\label{fig_cross}%
\end{figure}

Generally, the joint moment $\left\langle b_{m}b_{n}^{\ast}\right\rangle $
approaches $Y_{m,n}^{\pm}(r)$ only at long enough ranges where the factors
$\exp\left( - \frac{k^2}{2} \left\langle \left(  X_{m}^{\pm}\right)  ^{2}\right\rangle
\right)  $ become small. But Eq. (\ref{amn}) suggests that there are modes
whose moments $\left\langle b_{m}b_{n}^{\ast}\right\rangle $ are close to
$Y_{m,n}^{\pm}(r)$ at any $r$. These are the modes with
\begin{equation}
2kG_{m}\simeq\pi+2\pi J, \label{GmJ}%
\end{equation}
where $J$ is an integer. Indeed, for modes $m$ and $n$ satisfying this
condition, $\sin\left(  k\left(  G_{m}+G_{n}\right)  \right)  \simeq0$ and
$\cos\left(  k\left(  G_{m}-G_{n}\right)  \right)  \simeq1$. Then from Eqs.
(\ref{am2}) and (\ref{am0}), it follows that
\begin{equation}
\left\langle b_{m}b_{n}^{\ast}\right\rangle \simeq e^{-\frac{k^{2}}%
{2}\left\langle \left(  X_{m}-X_{n}\right)  ^{2}\right\rangle }\simeq
Y_{m,n}^{\pm}(r). \label{bmn}%
\end{equation}
The lower panel of Fig. 9 present normalized joint moments of
such modes. It is seen that these moments $\left\vert \left\langle b_{m}%
b_{n}^{\ast}\right\rangle \right\vert $ are reasonably well approximated by
the corresponding correlation functions $Y_{mn}^{\pm}(r)$.

Notice that in the case of adiabatic perturbation, formula (\ref{bmn}) is
applicable for all the modes. Indeed, the adiabaticity of $\delta c(r,z)$
requires that $\Delta_{\zeta}\gg D_{m}.$ If this condition is met 
(in our theory and numeric example we deal with  the inverse inequality), 
then $X_{m}^{+}(r)=X_{m}^{-}(r)$ and
Eq. (\ref{am-ray}) translates to

\begin{equation}
a_{m}(r)=\varphi_{m}\left(  z_{0}\right)  e^{ikX_{m}(r)}. \label{am-adiab}%
\end{equation}
Equation (\ref{bmn}) follows immediately from this formula. Thus, it turns out
that even though the adiabatic approximation in our example is not applicable,
the normalized cross-mode correlations $\left\langle b_{m}b_{n}^{\ast
}\right\rangle $ for modes satisfying condition (\ref{GmJ}) are properly
described using a simple adiabatic formula. It is worthwhile
to note, that the starting intensities of modes satisfying condition
(\ref{GmJ}) are $\left\vert a_{m}\left(  0\right)  \right\vert ^{2}%
\simeq2Q_{m}^{2}$ and their mean intensities weakly vary with range (see the preceding subsection), that is, they behave like the intensities
of adiabatic modes.

\section{Conclusion} \label{sec:conclusion}

In this paper the predictions of our ray-based analytic approach are compared
with results of full wave numerical simulation. Figures 3--7 and 9 present results of this comparison which are typical for most modes with numbers $m \leq 66 $. The comparison has confirmed the applicability of our approach
for the analysis of mode coupling in a deep water waveguide with sound speed
fluctuations induced by random internal waves. At a frequency of 100 Hz it can be used at ranges of a few hundred kilometers. However, for some
modes, especially for those which are weakly excited by the source and have
small initial amplitudes, the coincidence between the ray-based and pe-based
estmates may be much worse (not shown).

Since our approach is based on the WKB approximation, even at high frequencies
it cannot be used for those modes whose turning points are located in the
vicinity of the source depth or in the water bulk near the surface or bottom.
We have avoided these problems by setting the source at the sound channel axis
and restricting our attention to modes with turning points located well below
the surface and above the bottom. The WKB approximation and, hence, our
approach can be applied for modes with turning points on the waveguide
boundaries. But in the present paper such modes were not considered.

Numerical simulation has confirmed the prediction of Ref.
\cite{V89} that the range-dependencies of mode intensities and
other statistical moments are not smooth. They manifest jump-like
changes at ranges corresponding to upper turning points of the
mode rays.

Important advantage of our approach is its applicability for
evaluating the cross-correlations of mode amplitudes at different
frequencies needed for treating pulse propagation. It is clear
that analytical estimates for mode amplitudes at different
frequencies are readily derived along the same lines
as estimates for statistical moments given by Eqs. (\ref{am-mean}%
)--(\ref{am4-mean}). But this issue goes beyond the scope of the present work
and it is not broached here.

It should be emphasized that formula (\ref{am-ray}) is derived under
assumption that in the presence of perturbation the ray paths do not deviate
from their unperturbed positions. This assumption is valid only at short
enough ranges. In subsection V.B it is shown that the mode coupling
causes the equalization of mean mode intensities. According to Eqs.
(\ref{am2-mean}) and (\ref{am2}) mean intensities $\left\langle \left\vert
a_{m}(r)\right\vert ^{2}\right\rangle $ of modes with numbers close to $m$
approach $2Q_{m}^{2}$. Numerical simulation confirms this prediction (see Fig.
5). However, our approach cannot describe subsequent
changes of mode intensities with distance. In particular, it cannot be used to
study establishing the equipartition of energy among the modes in the limit
$r\rightarrow\infty$ predicted in Ref. \cite{DT1}.

In Sec. \ref{sec:master}, a numerical result of Refs.
\cite{Colosi2009,Colosi2012,Colosi2014} that the master equations properly
describe smoothed range-dependencies of mean mode intensities is explained
from the view point of our ray-based approach. It is shown that formula
(\ref{am2}) for the mean mode intensity obtained by smoothing the
range-dependent parameters of Eq. (\ref{am2-mean}) over the cycle of the mode
ray gives an approximate solution of the master equations valid in the high
frequency limit.

We assume that the expressions for $\left\langle a_{m}a_{n}^{\ast
}\right\rangle $ given by Eq. (\ref{amn-mean}) represent an
approximate solution of the complete system of $M^{2}$ equations
for these joint moments derived in the Markov approximation in
Refs. \cite{Creamer1996,Colosi2009}. However, for now, this
assumption has not been verified by direct substitution of Eqs.
(\ref{amn-mean}) in this system.

In Refs. \cite{Colosi2012,Colosi2014}, it is shown for a shallow
water waveguide, that the analytic expression for the joint moment
$\left\langle b_{m}b_{n}^{\ast}\right\rangle $ with $m\neq n$
derived in the adiabatic approximation may be valid if the sound
speed fluctuations are non-adiabatic. In the present paper, this
issue has not been studied in detail. However, we hope that our
comment on applicability of the adiabatic results in a
non-adiabatic environment made at the end of subsection
V.C may contribute to understanding this result of
Refs. \cite{Colosi2012,Colosi2014}.

Finally, note that Eq. (\ref{am-ray}) for the mode amplitude and the
expressions for statistical moments following from this formula can be easily
generalized to the case of a range-dependent unperturbed waveguide. This can
be done using analytical relations expressing mode amplitudes in a
range-dependent waveguide through parameters of ray paths
\cite{V99c,V2005a,Vbook2010}.

\section*{Acknowledgment}

The work was parially supported by the Program ``Fundamentals of acoustic
diagnostics of artificial and natural media'' of Physical Sciences
Division of Russian Academy of Sciences, Grants No. 13-02-00932
and 13-02-97082 from the Russian Foundation for Basic Research,
and Leading Scientific Schools grant N 339.2014.2.

\end{document}